\documentclass[twocolumn,trackchanges,twocolappendix]{aastex631}

\usepackage{savesym}
\savesymbol{tablenum}
\usepackage{siunitx}
\restoresymbol{SIX}{tablenum}

\begin{document}

\title{Reliability of {\it in-band} and broadband spectral index measurement: systematic study of the effect of signal to noise for uGMRT data}

\author[0009-0007-0120-5728]{Md Rashid}
\affiliation{Joint Astronomy Programme, Indian Institute of Science, Bangalore 560012, India}
\affiliation{Department of Physics, Indian Institute of Science, Bangalore 560012, India}
\author{Nirupam Roy}
\affiliation{Department of Physics, Indian Institute of Science, Bangalore 560012, India}
\author{J. D. Pandian}
\affiliation{Department of Earth and Space Sciences, Indian Institute of Space Science and Technology, Trivandrum 695547, India}
\author{Prasun Dutta}
\affiliation{Department of Physics, IIT (BHU), Varanasi 221005, India}

\author{R. Dokara}
\affiliation{Max-Planck-Institut für Radioastronomie, Auf dem Hügel 69, D-53121 Bonn, Germany}
\author{S. Vig}
\affiliation{Department of Earth and Space Sciences, Indian Institute of Space Science and Technology, Trivandrum 695547, India}
\author{K. M. Menten}
\affiliation{Max-Planck-Institut für Radioastronomie, Auf dem Hügel 69, D-53121 Bonn, Germany}

\begin{abstract}
Low radio frequency spectral index measurements are a powerful tool to distinguish between different emission mechanisms and, in turn, to understand the nature of the sources. Besides the standard method of estimating the ``broadband" spectral index of sources from observations in two different frequency ``bands", if the observations were made with large instantaneous bandwidth, the ``in-band" spectral index can be determined, either using images of emission at multiple frequency ranges within a band or using the novel Multi Term-Multi Frequency Synthesis (MT-MFS) imaging algorithm. Here, using simulated upgraded Giant Metrewave Radio Telescope (uGMRT) data, we have systematically studied the reliability of various methods of spectral index estimation for sources with a wide range of signal-to-noise ratio (SNR). It is found that, for synthetic uGMRT point source data, the MT-MFS imaging algorithm produces in-band spectral indices for SNR~$\lesssim100$ that have errors $\gtrsim 0.2$, making them unreliable. However, at a similar SNR, the sub-band splitting method produces errors $\lesssim 0.2$, which are more accurate and unbiased in-band spectral indices. The broadband spectral indices produce errors $\lesssim 0.2$ even for SNR $\gtrsim 15$, and hence, they are most reliable if there are no higher-order variations in the spectral index. These results may be used to improve the uGMRT observation and data analysis strategies depending on the brightness of the target source. 
\end{abstract}


\section{Introduction} \label{sec:intro}
The two distinct emission mechanisms that mostly contribute to the observed low radio frequency continuum are non-thermal synchrotron (e.g., from supernova remnants and active galactic nuclei) and the thermal bremsstrahlung (e.g., from H~{\sc ii} regions) radiation \citep{Wilson2013}. More often than not, astrophysical sources may have a mix of these two, e.g., in star-forming regions \citep{hunt2004, garay96}, Wolf-Rayet binaries \citep{setia2001}, the Galactic center region \citep{Law2008}, novae \citep{chomiuk2021, eyeres2006} and sometimes even H~{\sc ii} regions \citep{meng2019,padovani2019, 2020ApJ...898..172D}. The spectral energy distribution (SED), often characterized by the spectral index $\alpha$ (where specific intensity, $I_\nu\thicksim \nu^{\alpha}$ with frequency $\nu$), is a very effective tool to distinguish different emission mechanisms. For example, thermal emission is expected to have either $\alpha$ close to zero (i.e., a flat spectrum) or $\alpha \approx +2$ in the optically thin and thick regimes, respectively \citep{rlch5}. Optically thin synchrotron emissions typically have a spectral index in the range $-0.1 \geq \alpha \geq -2$ \citep{rlch6}; generally, a spectral index of $\alpha\le-0.5$ is taken to be the signature of non-thermal emission \citep{1999ApJ...527..154K}. On the other hand, synchrotron self-absorption, which happens in the case of optically thick radio synchrotron emission, has $\alpha$ close to $+2.5$.
\par
For extended sources, the spectral index ``map'' likely shows spatial variations of the SED. A spectral index map may be produced by fitting a power-law to continuum images at two or more frequency bands on a \textit{pixel-by-pixel} basis. This is referred to as the broadband spectral index image. However, there are a few caveats in determining the broadband spectral index. If the range of frequency is narrow, it is assumed that the emission can be modeled with a single power-law without any break or turnover.
\par
Technical upgrades of radio telescopes leading to the upgraded Giant Metrewave Radio Telescope \citep[uGMRT;][]{gupta17} and the Karl G. Jansky Very Large Array  \citep[JVLA;][]{EVLA2009}, along with new generation wide-band radio telescopes like the LOw Frequency Array \citep[LOFAR;][]{lofar2009} and the Australian Square Kilometer Array Pathfinder \citep[ASKAP;][]{askap}, enable us to obtain radio interferometric data that is contiguous over a wide range of frequency. The wide-band data obtained with these instruments result in high sensitivity and high dynamic range images due to the excellent \textit{uv}-coverage. The uGMRT provides up to 400 MHz instantaneous BW over the frequency range from 50 to 1500 MHz. Some of these frequency ranges are not covered by any other radio interferometers presently. Data from multiple GMRT bands have been used routinely to infer the broadband spectral index \citep[e.g.][]{ruta2012, 2018MNRAS.474.3808V, 2016MNRAS.456.2425V, stroe2013discovery, 2022A&A...664A.140K, 2019ApJ...884L..49P}. Recently, there have been attempts to produce the in-band spectral index images determined by fitting a polynomial to sub-band images for some low-frequency telescopes \citep[e.g.][]{Venturi2022}, including the studies carried out using the uGMRT \citep[e.g.][]{10.1093/mnras/sty3032}. However, despite the advantages of instantaneous large BW of observations, in-band spectral index studies with the uGMRT have been limited.
In this regard, the Multi-Frequency Synthesis (MFS) algorithm, in addition to improving the effective \textit{uv}-coverage, includes spectral index variations across the source for better image reconstruction \citep{1990MNRAS.246..490C}. The Multi Term-MFS (MT-MFS) imaging algorithm uses a Taylor series expansion of the model flux density ($F^\mathrm{sky}_{\nu}$) around a reference value $\nu_{0}$:
\begin{equation}
    F^\mathrm{sky}_{\nu}=\sum_{i} F^\mathrm{sky}_{i} \left( {\frac{\nu - \nu_0}{\nu_0}} \right)^{i}
	\label{eq:mfs_poly}
\end{equation}%
to reconstruct the spectral structure, assuming a power-law scaling:
\begin{equation}
    F^\mathrm{sky}_{\nu}=F^\mathrm{sky}_{\nu_0} \left( {\frac{\nu}{\nu_0}} \right)^{\alpha + \beta \log(\nu / \nu_0) }
    \label{eq:pl_scaling}
\end{equation}%
where, $\alpha$ is the spectral index \citep{rau2016}. Any higher-order variations in the SED can also be modeled by the curvature ($\beta$) parameter using higher-order Taylor terms if a sufficient number of flux density measurements across different frequencies are available. In principle, with sufficient fractional bandwidth ($\Delta\nu/\nu_0$), one can determine $\alpha$ and $\beta$, depending on the number of terms retained in the Taylor expansion. Using synthetic EVLA data \citet{rau2011} showed that the Multi Scale-MFS imaging algorithm could reconstruct the spectral structure better compared to conventional imaging. Along with that, they pointed out that the uncertainty in the measured spectral index increases at lower signal-to-noise ratios (SNR), but no quantitative limits were given. \citet{rau2016} also tested the MT-MFS algorithm in combination with wide-band AW-projection \citep{2013ApJ...770...91B} on a field crowded by point sources with a noiseless simulation of VLA data. However, the imaging process produced small numerical noise in the image plane. They found that the uncertainty in the spectral index values for lower SNRs deteriorated more rapidly than in the intensity, even when the noise was predominantly numerical. For the weakest sources (SNR $\thicksim 30$), they could not even estimate spectral indices. However, these results are specific to VLA, and a systematic investigation of spectral index uncertainties with MT-MFS and other existing methods (particularly for low-frequency telescopes such as uGMRT) is indispensable.
\par
Reliable radio spectral index estimation depends on many factors: SNR, \textit{uv}-coverage, wide-field effects, the frequency dependence of the primary beam structure, fractional bandwidth, the angular extent of the source, etc. In this work (Paper-I), a systematic investigation of the reliability of the spectral index recovered from different methods at different SNRs is presented for low-frequency synthetic uGMRT data. The work is motivated by the detection of non-thermal emission in the deep continuum image of the H~{\sc ii} region associated with the Orion Nebula, as emphasized in our ongoing study (Rashid et al., in prep.; Paper-II) using the uGMRT. Before interpreting the results (particularly those based on the spectral index image), it is imperative to verify their reliability thoroughly, as only a handful of H~{\sc ii} regions are known to have both thermal and (some) non-thermal emissions \citep[e.g.,][] {meng2019,padovani2019,2023A&A...670A...9D}. 
\begin{table}
	\centering
	\caption{Observational parameters for simulating visibility data sets.}
	\label{tab:tel_parm}
	\begin{tabular}{lr} 
		\hline
		Telescope & uGMRT\\
		Reference time (UTC) & 2017/01/09/12:27:23\\
		  Obs. phase center & $05^{\rm h}35^{\rm m}17\rlap{.}^{\rm s}5$, $-05^\circ23'36''$\\
		No. of antenna & 30 \& 16 \\
		Reference frequency (MHz) & 400 \& 650 \\
		Channel width, $\delta\nu$ (kHz) & 390.625\\
		Band Width, $\Delta\nu$ (MHz) & 200\\
		No. of channels & 512\\
		Stokes & RR $\&$ LL\\
		Integration time, $\delta t$ (s) & 10\\
		On-source time, $\tau$ (Hr) & 6\\
		\hline
	\end{tabular}
\end{table}
We have used simulated uGMRT visibility data to test the recovery of the in-band and broadband spectral indices for point sources with varying SNR, \textit{uv}-sampling, and different input spectral indices. Both MT-MFS spectral images and sub-band spectral fitting methods have been used for the in-band spectral analysis; we have carried out a conventional pixel-by-pixel spectral fitting to estimate broadband spectral index from two different bands. The primary goal of the current work is to determine the best method to obtain reliable values of the spectral index for uGMRT images with low SNRs, as well as to understand the advantages and limitations of the various methods employed for low-frequency data. In a follow-up paper (Paper-II), we will present a systematic study of the additional complications of spectral index estimation for extended emission, focusing on the effect of fractional bandwidth and wide-field imaging effects like variations of \textit{uv}-coverage and beam width. In particular, we will study the implications of these for low-frequency data and the interpretation of the spectral index study of the Orion nebula with the uGMRT. The rest of the paper is organized as follows: In \S \ref{sec:s2}, we discuss the simulation, imaging, and different in-band and broadband spectral indices estimation strategies. \S \ref{sec:res} lists the important results and outcomes of the analysis; in \S \ref{sec:disc}, these are discussed and interpreted, and a summary and conclusion are presented in \S \ref{sec:conc}. Throughout the paper, we have referred to the continuum or wideband SNR as SNR unless stated otherwise.

\section{Simulations, imaging, and spectral analysis}
\label{sec:s2}
The visibility data sets of point sources (with spectral indices -2, 0, 2) at the phase center with uGMRT baseline configuration and telescope parameters were simulated using tasks and tools from the Common Astronomy Software Applications  \citep[CASA;][]{2007ASPC..376..127M}\footnote{\url{https://casa.nrao.edu/}} version 6.4. Table \ref{tab:tel_parm} lists the observational parameters used for the simulations. Different imaging and spectral index estimation strategies, described below, are used to infer the reliability of the spectral index measurements from various methods.

\subsection{In-band spectral index estimation}
\label{sec:s2.1}

\subsubsection{MT-MFS}

The uGMRT band-3 (250--500 MHz) configuration has been chosen to simulate the visibilities corresponding to three different spectral regimes, namely optically thin non-thermal, optically thin thermal and optically thick thermal emission, respectively. The  phase center is taken in the direction of the Orion nebula at $(\alpha, \delta)_{{\rm J}2000}$ = $05^{\rm h}35^{\rm m}17\rlap{.}^{\rm s}5$, $-05^\circ23'36''$. Data are generated for both ``RR" and ``LL" correlations, with a total bandwidth of 200 MHz divided into 512 channels, 10~s integration time, and a total on-source time ($\tau$) of six hours. For the simulated image cube containing the point source with input flux density and spectral index, the task \texttt{tclean} in CASA is used to predict the model visibilities. Next, the task \texttt{simulator} is used to add Gaussian random noise using the \texttt{simplenoise} mode of \texttt{setnoise} that allows introducing frequency-dependent noise per visibility
\begin{equation}
  \sigma_\mathrm{vis}=\frac{T_\mathrm{sys}}{G} \cdot \frac{1}{\sqrt{\delta t \cdot \delta \nu}}.
  \label{sigma_vis}
\end{equation}
for the choice of frequency resolution ($\delta \nu$), integration time ($\delta t$), system temperature ($T_{\rm sys}$), and antenna gain ($G$). In practice, the on-source time to achieve a target RMS noise is longer than the theoretical estimate due to various factors, including RFI, overhead time, baseline coverage, etc.; these effects are absorbed in the ``fudge factor'' ($f$)\footnote{\url{http://www.ncra.tifr.res.in:8081/~secr-ops/etc/etc_help.pdf}}, which is the ratio of the actual to the theoretical on-source time. For band-3, $f$ lies between 3 and 5. Hence, we have scaled $\sigma_\mathrm{vis}$ with $f = 4$ to achieve a realistic noise in the final image. The image noise ($\sigma_\mathrm{im}$) scales with $\sigma_\mathrm{vis}$ as
\begin{equation}
    \sigma_\mathrm{im}= \frac{\left(\sigma_\mathrm{vis}/\sqrt{f}\right)}{ \sqrt{n_\mathrm{ch}}  \times \sqrt{n_\mathrm{bl}} \times \sqrt{n_\mathrm{pol}} \times \sqrt{n_\mathrm{int}} },
	\label{eq:sigma_image}
\end{equation}%
where $n_\mathrm{int} = \tau/\delta t$ is the number of integrations, and $n_\mathrm{ch}$, $n_\mathrm{bl}$, and $n_\mathrm{pol}$ are the number of channels, baselines, and polarisation, respectively. From our noise-free simulation with the full synthesis configuration, a dynamic range of $\thicksim 10^7$ is obtained in the continuum image. To achieve different SNRs while keeping the \textit{uv}-coverage similar, the input flux densities of the sources were varied instead of the on-source times. Input point source flux densities of 0.25, 0.5, 1, 1.5, 2, and 2.5 mJy were used to achieve SNRs of 25, 50, 100, 150, 200, and 250, respectively. 
\par
The CASA task \texttt{tclean} is used to make images of size $\thicksim \SI{6}{\arcminute} \times \SI{6}{\arcminute}$, from the simulated visibility data using the MT-MFS deconvolver with two Taylor terms. For the Briggs weighting scheme \citep{briggs95} with a fiducial robust parameter of 0, $\sigma_\mathrm{vis}$ = 0.25 Jy results in an image noise of $\thicksim 10$ $\mu$Jy~beam$^{-1}$ and the corresponding synthesised beam of $\thicksim \SI{12}{\arcsecond} \times \SI{6}{\arcsecond}$. With 435 baselines, two polarizations, and 200 MHz bandwidth, this is the typical noise for a realistic observation with the uGMRT.
\par
The use of MT-MFS imaging in the \texttt{tclean} task produces Taylor coefficient maps as well as $\alpha$ and  $\beta$ maps by taking the ratio of these coefficient maps. The spectral indices are calculated up to one-tenth of the peak residual values during deconvolution by the algorithm. In addition, a spectral index error map is also produced  \citep[see][for the mathematical expression of the error]{rau2011}. In this work, the spectral index distribution maps have been produced from the simulated data at different SNRs using two Taylor terms in the MT-MFS imaging algorithm.

\subsubsection{Sub-band Splitting}
The in-band spectral index is also estimated by sub-band imaging of the synthetic visibility data. The 200 MHz bandwidth data were split into four frequency chunks of 50 MHz each using the task \texttt{split} and imaged separately with \texttt{tclean} parameters as described above. It is to be noted that the SNR in each sub-band image is a factor of two lower compared to that of the MT-MFS image using the entire bandwidth.
\par
To determine the spectral indices from the sub-band images, all the sub-band images were first smoothed to a common resolution using the \texttt{imsmooth} task in CASA. Then, the point sources in the individual sub-band images were fitted with a Gaussian component using the task \texttt{imfit}. These recovered flux densities, along with their uncertainties, were then fitted with a power-law to obtain the spectral index and its error using the standard $\mathbf{\chi}^2$ minimization function \texttt{curvefit} from the scipy optimization module in python. All these steps were performed for different SNRs for all three values of the fiducial spectral indices.

\subsection{Broadband spectral index estimation}
\begin{figure*}
    \centering
    \includegraphics[trim={0 0 0 1cm},clip,width=8.0cm]{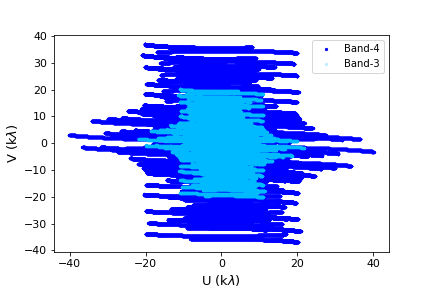}
    \includegraphics[trim={0 0 0 1cm},clip,width=8.0cm]{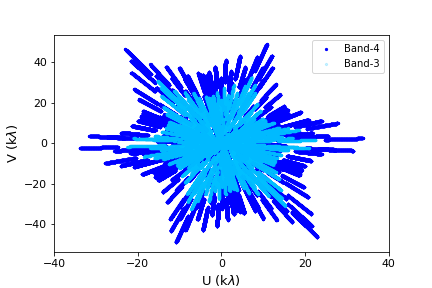}
    \caption{Comparison of \textit{uv}-coverage of simulated band-3 and band-4 data. Band-4 data is plotted in dark blue, and band-3 data is plotted in transparent light blue color, both in wavelength units. Left: \textit{uv}-tracks of 6 hours on-source time at a single frequency. Right: snapshot (single time stamp) of \textit{uv}-tracks with 512 channels.}
    \label{fig:uv-cov}
\end{figure*}
For broadband spectral index analysis, we have chosen band-3 (250--500 MHz) and band-4 (550--850 MHz) of the uGMRT. For band-3, we have used the same data set we simulated for in-band spectral index analysis (see \S \ref{sec:s2.1}). To simulate band-4 data, we have extended the same SED used for the band-3 simulations ($\alpha_\mathrm{true}=$  $-2$, 0, 2) to obtain the band-4 flux densities. Given the different flux densities of the model point source, one has to adjust the observation times to achieve the same SNR in band-4 as in band-3. We have used 6 hours of on-source time for the band-3 observations to ensure adequate \textit{uv}-coverage. A comparison between the band-3 and 4 \textit{uv}-coverage is shown in Figure \ref{fig:uv-cov}, showing how the distribution of \textit{uv} data points varies with time and bandwidth. In order to avoid complications related to changing uv-coverage and/or observations spread over multiple days, we have obtained the desired SNR by scaling $\sigma_\mathrm{vis}$ with the appropriate factor rather than increasing the on-source time.
\par 
We used the same pixel and image sizes for imaging the data in both bands. A threshold of 5 times the estimated rms noise is used for deconvolution. 
The image rms noise obtained in band-3 is $\thicksim 10$ $\mu$Jy~beam$^{-1}$ for all spectral models  and for band-4 the noise levels are $\thicksim 4, 10$ and $28$ $\mu$Jy~beam$^{-1}$ for $\alpha_\mathrm{true}=-2,0$ and 2, respectively. 
\par
To estimate the broadband spectral index from the band-3 and band-4 images, they were first smoothed to the same angular resolution using the \texttt{imsmooth} task before making the spectral index map. Taking flux density $F_\nu \propto \nu^{\alpha}$, the spectral index $\alpha$ can be written as
\begin{equation}
    \alpha = \frac{\log(F_{1}) - \log(F_{2})}{\log(\nu_{1}) - \log(\nu_{2})}
    \label{eq:bb-alpha}
\end{equation} %
where $F_{1}$ and $F_{2}$ are flux densities at reference frequencies $\nu_{1}$ $\&$ $\nu_2$, respectively. The uncertainty in $\alpha$ is 
\begin{equation}
    \sigma_\mathrm{\alpha} = \frac{1}{|\log(\nu_1/\nu_2)|} \cdot \sqrt{\left(\frac{\sigma_\mathrm{F_{1}}}{F_1}\right)^2 + \left(\frac{\sigma_\mathrm{F_{2}}}{F_2}\right)^2},
    \label{eq:bb-error}
\end{equation} %
where $\sigma_\mathrm{F_1}$ and $\sigma_\mathrm{F_2}$ are the rms uncertainties of the flux maps. We have used the \texttt{immath} task in CASA to evaluate these expressions using the continuum images and 
to produce the $\alpha$ and $\sigma_\mathrm{\alpha}$ maps. We have used a flux cutoff of 5 times the rms noise of the continuum image for both bands.

\section{Results}
\label{sec:res}
\subsection{In-band spectral index}
The results discussed in this work are from a sample noise realization, an extreme case where the retrieved spectral index $\alpha$ deviates the most from the input spectral index $\alpha_\mathrm{true}$. To examine the variation in the recovered spectral index across different noise realizations, we conducted simulations for 32 random noise realizations using an SNR of 25 and 100, with $\alpha_\mathrm{true} = 0$. However, unless explicitly stated otherwise, the results and discussions are primarily based on the analysis of one specific noise realization (the most extreme case). Furthermore, we deemed $\pm 0.2$ deviation on $\alpha$ acceptable, as it offers sufficient confidence in distinguishing between various emission mechanisms and hence retrieved $\alpha$ meeting this criterion is referred to as reliable and vice versa.
\par
For all three values of $\alpha_\mathrm{true}$, the retrieved MTMFS $\alpha$ have high positive biases (up to 0.8 for SNR = 25; see Table \ref{table:all-method_alpha}) for this noise realization. The bias is defined as $\left|\alpha - \alpha_\mathrm{true}\right|$. The bias value decreases as the SNR increases, but they are different for different spectral regimes. At an SNR of 100 or greater, the total error on the spectral index, $\Delta \alpha$ ($=\left|\alpha - \alpha_\mathrm{true}\right|+\left|\sigma_\mathrm{\alpha}\right|$) at the central pixel is $\lesssim 0.2$. Ideally, the spectral index map for the point source should be correlated and must have the same value across the synthesized beam. However, our analyses show that even for a simple point source at the phase center, the spectral index map produced by the MT-MFS algorithm shows a variation across the beam, contrary to our expectations.
\begin{table*}
\caption{Spectral index and associated errors from different methods for one realization. The retrieved spectral index value $\alpha$ and associated error $\sigma_\mathrm{\alpha}$ from three methods are listed for a particular noise realization at different SNR. The input spectral index $\alpha_\mathrm{true}$ is also indicated above the columns.}
	\begin{center}
		\begin{tabular}{c|cccccc|cccccc|cccccc}
	\hline		
			&\multicolumn{6}{c|}{MT-MFS} & \multicolumn{6}{c|}{Sub-band} & \multicolumn{6}{c}{Broadband}\\

	\cline{2-19}	
 SNR	&\multicolumn{2}{c}{$\alpha_\mathrm{true}=-2$} & \multicolumn{2}{c}{$\alpha_\mathrm{true}=0$} & \multicolumn{2}{c|}{$\alpha_\mathrm{true}=2$} & \multicolumn{2}{c}{$\alpha_\mathrm{true}=-2$} & \multicolumn{2}{c}{$\alpha_\mathrm{true}=0$} & \multicolumn{2}{c|}{$\alpha_\mathrm{true}=2$}&\multicolumn{2}{c}{$\alpha_\mathrm{true}=-2$} & \multicolumn{2}{c}{$\alpha_\mathrm{true}=0$} & \multicolumn{2}{c}{$\alpha_\mathrm{true}=2$}\\
			\cline{2-19}
			  & $\alpha$ & $\sigma_\mathrm{\alpha}$ & $\alpha$ & $\sigma_\mathrm{\alpha}$ & $\alpha$ & $\sigma_\mathrm{\alpha}$& $\alpha$ & $\sigma_\mathrm{\alpha}$ & $\alpha$ & $\sigma_\mathrm{\alpha}$ & $\alpha$ & $\sigma_\mathrm{\alpha}$& $\alpha$ & $\sigma_\mathrm{\alpha}$ & $\alpha$ & $\sigma_\mathrm{\alpha}$ & $\alpha$ & $\sigma_\mathrm{\alpha}$\\
	\hline		
   \hline
			15 & - & - & - & - & - & - & - & - & -  & - & - & - & -2.11 & 0.17 & -0.04 & 0.17 & 1.93 & 0.17 \\
			25 & -1.21 & 0.17 & 0.63 & 0.16 & 2.4 & 0.41 & -1.61 & 0.6 &  -0.02 & 0.48 & 1.72 & 0.52 & -2.1 & 0.1 & -0.03 & 0.1 & 1.95 &  0.11 \\  
			50 & -1.58 & 0.13 & 0.33 & 0.05 & 2.19 & 0.2 & -1.9 & 0.24 & -0.01 & 0.25 & 1.86 & 0.27 & -2.09 & 0.06 & -0.01 & 0.05 & 1.96 & 0.06 \\
		    100 & -1.77 & 0.08 & 0.17 & 0.02 & 2.09 & 0.1 & -1.96 & 0.12 & 0 & 0.12 & 1.93 & 0.14 & -2.09 & 0.03 & -0.01 & 0.03 & 1.97 & 0.03 \\
			150 & -1.84 & 0.05 & 0.11 & 0.01 & 2.05 & 0.06 & -1.97 & 0.08 & 0 & 0.08 & 1.95 & 0.09 & -2.08 & 0.02 & 0 & 0.02 & 1.97 & 0.02 \\
			200 & -1.87 & 0.04 & 0.08 & 0.01 & 2.03 & 0.05 & -1.98 & 0.06 & 0 & 0.06 & 1.96 & 0.07 & - & - & - & - & - & -\\
			250 & -1.9 & 0.03 & 0.07 & 0.01 & 2.02 & 0.04 & -1.99 & 0.05 & 0 & 0.05 & 1.97 & 0.05& - & - & - & - & - & -\\
			\hline
		\end{tabular}
	\end{center}
	
	\label{table:all-method_alpha}
\end{table*}
\par
The values of $\alpha$ obtained from power-law fitting to point source flux densities obtained from sub-band images are listed in Table \ref{table:all-method_alpha} along with the fit errors. It can be seen that the spectral index values estimated from the sub-band method have no significant bias for $\alpha_\mathrm{true}=0$. However, the fit error is $\pm 0.5$ for SNR 25 and $\pm 0.25$ for SNR 50. Hence, the fitting yields acceptable $\alpha$ values for SNRs $> 50$.  Interestingly, the values are biased towards positive values for $\alpha_\mathrm{true}=-2$ at lower SNRs, whereas for $\alpha_\mathrm{true}=2$, the values are negatively biased for this realization. However, the fitted $\alpha$ are acceptable above SNR 50 as the bias, and the fit uncertainty diminishes significantly (e.g., from $\pm 0.6$ to 0.12 for $\alpha_\mathrm{true}=0$) and $\Delta\alpha$ lies in the acceptable range for all the $\alpha_\mathrm{true}$.  

\subsection{Broadband spectral index}
In these measurements, the retrieved $\alpha$ at the central pixels are consistent with $\alpha_\mathrm{true}=0$ even at SNRs 15. The error $\Delta \alpha <0.2$ for all the input spectral index down to SNR 15. However, the $\alpha$ values are slightly biased towards the negative values. The bias is highest ($\thicksim 0.1$) for $\alpha_\mathrm{true}=-2$. Table \ref{table:all-method_alpha} lists all the retrieved values for the spectral indices. The values are trustworthy at SNR levels as low as 15.
\par
We have also done point source fitting on the images of both bands and found that the peak flux density differs from the integrated flux density by $\lesssim 1\%$ ($2.5\%$ maximum for SNR 15) most of the time. If we do a power law fitting with the integrated flux density, this could result in an additional error on $\alpha$ by $\lesssim0.06$ (1.7 maximum for SNR=15). Hence, the effect of pixelation on spectral index measurement is not dominant in our pixel-by-pixel power law fitting.

\subsection{Multiple point source simulation}
 \begin{figure}
    \centering
    \includegraphics[trim={0 0.5cm 1cm 1cm},clip,width=\linewidth]{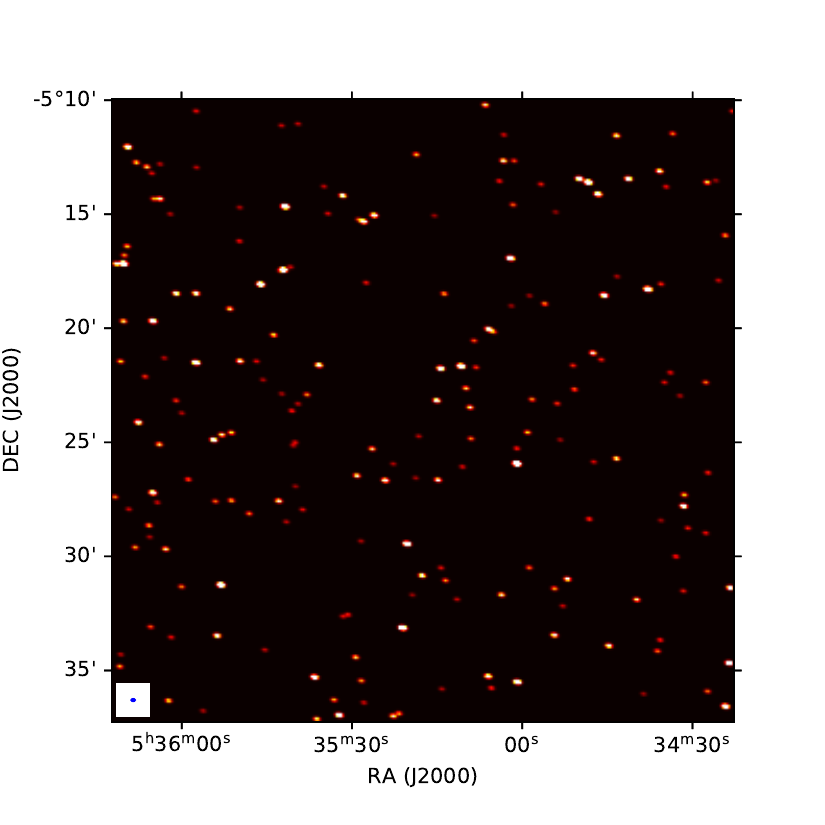}
    \caption{Input model for simulation of multiple point source following $dN/dS\propto S^{-1.6}$ relation. The image is convolved with a beam of $\thicksim \SI{12}{\arcsecond} \times \SI{6}{\arcsecond}$ (shown in the bottom left corner of the image) for ease of visibility.}
    \label{fig:model-multi-pts}
\end{figure}
 We have also simulated a field crowded with point sources within the 50\% of Half Power Beam Width (HPBW) of the uGMRT antenna to check how the in-band spectral index recovery by MT-MFS imaging is affected by the presence of multiple point sources distributed within a small region covered by the primary beam. We empirically used low radio frequency differential source count as a function of flux density relation measured from GMRT studies, $dN/dS \propto S^{-1.6}$ \citep{2016MNRAS.459..151C} to populate our model in a more realistic fashion. The source locations are distributed following a uniform random distribution within 50~\% of the HPBW. The input model is shown in Figure \ref{fig:model-multi-pts}. In this exercise, we have only used $\alpha_\mathrm{true}=0$. There are 199 sources between flux density of 100~$\mu$Jy-beam$^{-1}$ to 2~mJy-beam$^{-1}$ covering an SNR range of 10 to 200. 
 \begin{table}
    \caption{Number of sources at different SNR bins for multiple point source simulation (off source image noise $\sim10$~$\mu$Jy-beam$^{-1}$)}
    \centering
    
    \begin{tabular}{ccr}
        \hline
        \hline
        Bin no. &SNR Range & N \\
        \hline
        1& 10.0 -- 13.5 & 39 \\
        2& 13.5 -- 18.2 & 33 \\
        3& 24.6 -- 33.1 & 28 \\
        4& 33.1 -- 44.7 & 23 \\
        5& 44.7 -- 60.3 & 19 \\
        6& 60.3 -- 81.4 & 16 \\
        7& 81.4 -- 109.8 & 13 \\
        8&109.8 -- 148.2 & 11 \\
        9&148.2 -- 200.0 & 9 \\
        \hline
    \end{tabular}
    
    \label{tab:multi-snr}
\end{table}
 We have used a logarithmic binning scheme; the number of sources in different SNR bins is listed in Table \ref{tab:multi-snr}. However, quantifying the uncertainty will not be statistically significant due to the low number of sources in higher SNR bins. To address this problem, we simulated another data set with the same SNR range divided into ten linearly spaced bins and populated them with thirty-two sources in each bin. 
 \begin{figure}
    \centering
    \includegraphics[trim={0 0 1cm 1cm},clip,width=\linewidth]{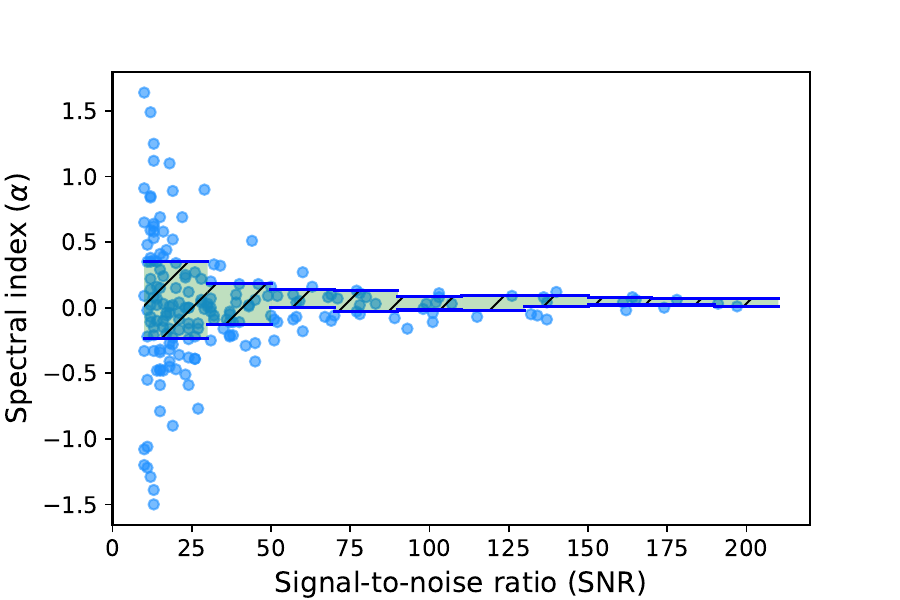}
    \caption{Retrieved $\alpha$ (from MT-MFS imaging) versus SNR from multiple point source simulation following $dN/dS\propto S^{-1.6}$ relation are plotted using transparent light blue points. The green shaded region shows IQR q$_{13}$ at different SNR bins from uniform binning case.}
    \label{fig:snr-dnds}
\end{figure}
 The MT-MFS spectral indices for the dN/dS scaled model are plotted in Figure \ref{fig:snr-dnds}. Since the input spectral index, $\alpha_\mathrm{true}=0$, the retrieved $\alpha$ values are the errors in the measured spectral index with respect to the input value for the individual sources.  The green-shaded regions represent the interquartile range (IQR) of quartiles 1 to 3 (q$_{13}$) obtained from different bins of the uniformly binned simulation. It is evident from this figure that the spread of the values is decreasing with increasing SNR. The findings from the multiple source simulation align well with the results obtained from single point source simulations for MT-MFS spectral index measurement. 
 
\section{Discussion}
\label{sec:disc}
\begin{figure*}
\centering
	\includegraphics[trim={0 0 0 1cm},clip,scale=0.5]{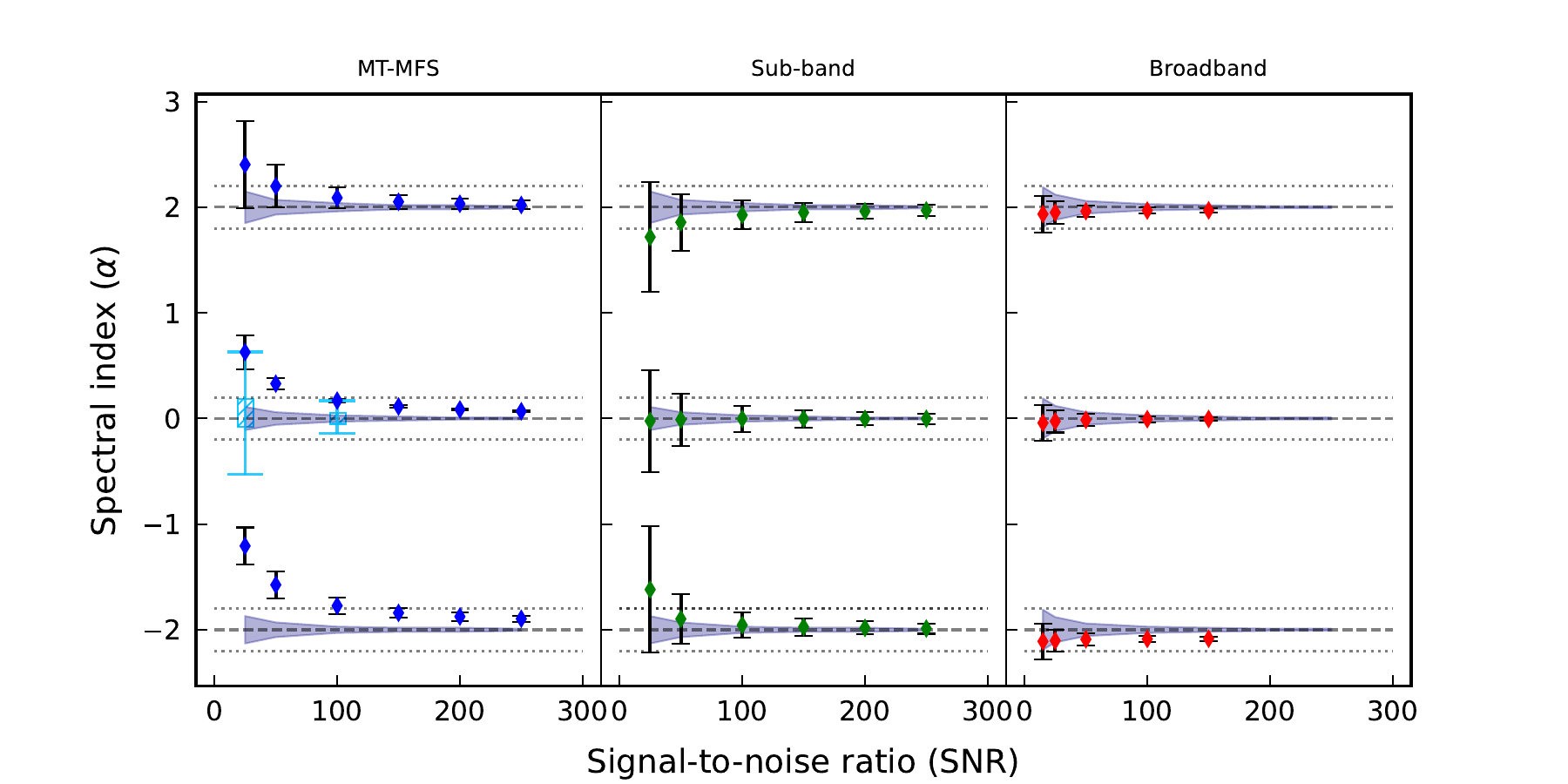}%
      \caption{Spectral index vs. SNR for different methods. Dashed lines represent input spectral index values $\alpha_\mathrm{true}$. Dotted lines around dashed line marks $\pm 0.2$. The blue-shaded region between them is the analytically calculated error for ``perfect" measurement. Blue markers are used for plotting the $\alpha$ obtained from the MT-MFS method, green markers are for the sub-band method, and red markers are for broadband spectral index measurement. Values are taken from central pixels of the  $\alpha$ maps for MT-MFS and broadband methods. The fitted $\alpha$ and fit error are directly plotted for the sub-band method. The shaded light blue box for the case of SNR 25 and 100 from $\alpha_{true}=0$ shows the IQR q$_{13}$ range from 32 realizations, and the extended bar with a cap on the box shows the entire range of retrieved $\alpha$ from 32 noise realization.}
        \label{fig:all-alpha-err}
\end{figure*}
Radio spectral index-based science from low-frequency radio telescopes such as LOFAR, ASKAP, MeerKAT, and uGMRT often requires sub-band and broadband spectral index maps. The reliability of spectral index maps is determined by calculating the uncertainties from error propagation, comparing with the existing literature or from statistical arguments for a large sample of targets \citep[e.g.][]{2016A&A...593A..86V, 2016MNRAS.457.4160H, 2016MNRAS.463.2997M, 2022ApJ...924...64I, 2022MNRAS.513.1300N, 2021MNRAS.508.3995B}. Most of these studies investigate the dependence of the spectral index on the SNR, but to the best of our knowledge, no systematic study examines the reliability of the values based on data from the low-frequency telescopes mentioned above.
\par
Since the MT-MFS algorithm is widely used to reconstruct radio continuum images, rigorous testing is necessary to establish the reliability of the determined spectral index. In this study, we have investigated how well the spectral index is being recovered for different SNRs with MT-MFS and other existing methods for the synthetic data in bands 3 and 4 of the uGMRT (250--500 and 550--850 MHz, respectively). Figure \ref{fig:all-alpha-err} shows the recovered $\alpha$ and associated errors from three methods. We find that for many noise realizations, the $\alpha$ values start to deviate significantly from the input spectral indices for SNRs lower than 100 and may be as high as $0.8 \geq \Delta \alpha \geq 0.1$ for SNR between 25 to 100. Also, the bias, $\left|\alpha-\alpha_\mathrm{true}\right|$, is more dominant in the negative spectral regime than in the flat and positive power-law regimes. However, for SNRs above 100, the $\alpha$ values could be useful in distinguishing thermal and non-thermal emission as the total error $\Delta \alpha$ ($=\left|\alpha - \alpha_\mathrm{true}\right|+\left|\sigma_\mathrm{\alpha}\right|$) $\leq 0.2$. Similarly, the retrieved $\alpha$ values are reliable at a similar SNR range for the multiple point source simulation as well (see Figure \ref{fig:snr-dnds}).

\begin{figure*}
    \centering
    \includegraphics[trim={0 1.6cm 0 2.5cm},clip,scale=0.55]{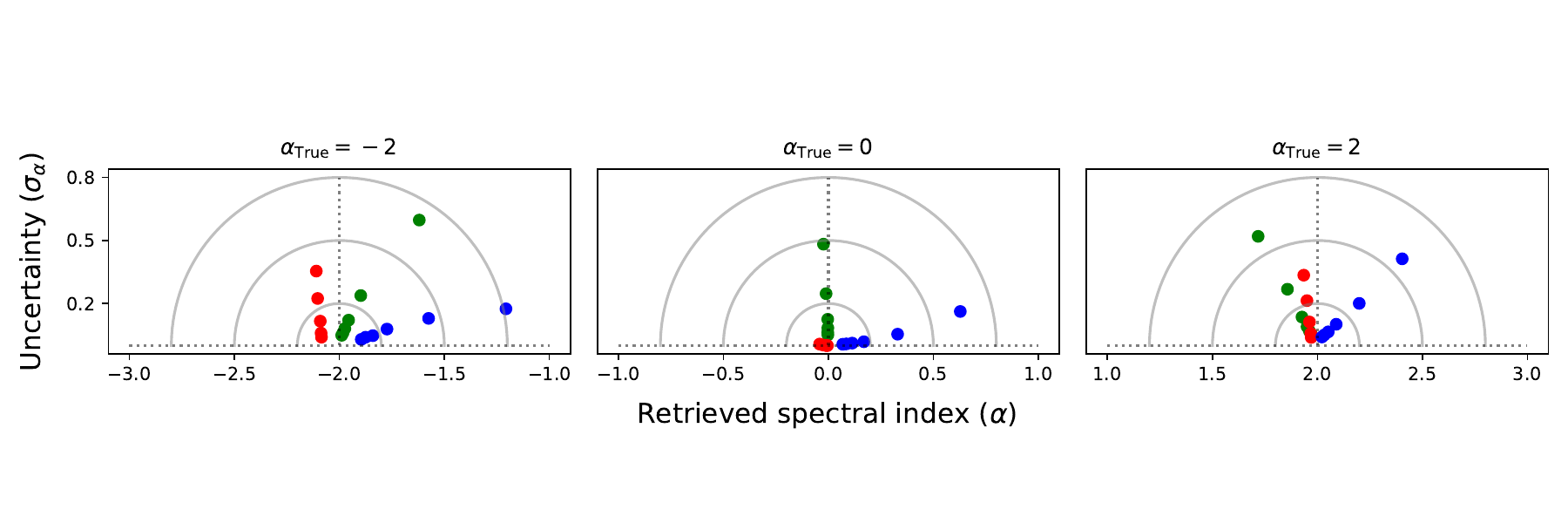} 
    \caption{Uncertainty vs. retrieved spectral indices plot for different input spectral indices. Blue, green, and red circles are plotted for MT-MFS, sub-band, and broadband data, respectively. Semi-circular contours are drawn at 0.2, 0.5 and 0.8 radii. Data points are in increasing order of signal-to-noise ratio with decreasing contour radius.}
    \label{fig:acc_prec}
\end{figure*}
\par
Our sub-band spectral analysis indicates that the inferred $\alpha$ is reliable above SNRs of 50, although there may be small biases, i.e., the differences between $\alpha_\mathrm{true}$ and $\alpha$, in the negative and positive spectral indices for low SNR cases. 
\par
Conventionally, the broadband method is used for making spectral index maps. Not surprisingly, in our analysis, we have found that it produces the most reliable spectral indices at SNRs as low as 15. Nevertheless, there are some caveats in using broadband spectral index maps. For images in different bands, there are always some flux calibration uncertainties in addition to the thermal noise. These uncertainties are typically $\thicksim10\%$ which can be incorporated in equation \ref{eq:bb-error} to give the final uncertainty in $\alpha$ values \citep{2014JKAS...47..195K}. Furthermore, without a continuous frequency coverage, it is challenging to incorporate higher-order variations of the spectral index, e.g., a spectral break, turnover, steepening, etc. In addition, the effects of different \textit{uv}-coverage and \textit{uv}-filtering at different scales can further bias the spectral index.
\par
Figure \ref{fig:all-alpha-err} shows a plot of spectral index vs. SNR comparing the performance of three methods based on bias and error values. All the results represented by diamond-shaped markers in the different panels correspond to one particular noise realization and the error bars are shown in black for it. With the uncertainties described by these, it seems that the estimated spectral index shows a positive bias for the MT-MFS method. The bias is particularly significant at lower SNRs. As mentioned earlier, we assess the true uncertainty in spectral index determination by repeating the spectral index estimation for 32 different noise realizations for $\alpha_\mathrm{true}=0$ and SNR = 25 and 100. The box-shaped shaded region in light blue color in the first panel shows the IQR $q_\mathrm{13}$ for these 32 noise realizations, while the extended bar with cap shows the entire range of retrieved $\alpha$. Clearly,  the error associated with the estimated $\alpha$ values is much smaller than the range of variation of $\alpha$ for the different noise realizations; only at SNR $\geq$ 100 the MT-MFS spectral index estimations are reliable. 
\par
From the second panel of Figure \ref{fig:all-alpha-err}, it is clear that the sub-band method performs better in terms of both bias and error. However, this method is also limited by the fact that at low SNR, the sensitivity of individual sub-band should be high enough and should have similar \textit{uv}-sampling. Finally, the third panel shows that the broadband method produces the most accurate spectral index consistent with $\alpha_\mathrm{true}$ within $2\sigma$ uncertainty for an SNR as low as 15.
\par
In Figure \ref{fig:acc_prec}, the uncertainties are plotted against the retrieved spectral indices. 
The accuracy and precision of the different methods in different spectral regimes, as well as any trend depending on the spectral regime, can be understood more clearly from this figure. As for the flat spectrum case, the broadband value is the most precise and accurate, but the sub-band spectral indices are less accurate, although they are precise to some extent. The values produced by MT-MFS are neither very accurate nor very precise. Similar arguments could be deduced for other spectral regimes as well. 
\par
These results are expected to help in deciding the observational and analysis strategies for both targeted observations and sky surveys with radio interferometer arrays, especially for the uGMRT users. In particular, if the {\it in-band} spectral index is of interest, one must target SNRs $\geq 50$ and rely more on the sub-band splitting method of spectral index estimation for SNR in the range of $50 - 100$.

\section{Concluion}
\label{sec:conc}
We have simulated the uGMRT low-frequency data for point source models at different SNRs and input spectral indices. We then processed these data to estimate spectral indices from these synthetic data by three different methods, i.e., MT-MFS, sub-band splitting, and broadband spectral fitting. We have assessed the reliability of the spectral index maps produced by these methods and draw the following main conclusions from this study:
    \begin{enumerate}
    
	\item The in-band spectral index maps from MT-MFS imaging are reliable for SNRs of 100 and higher; there may be significant bias (much higher than 0.2) for a single measurement and large uncertainties across different noise realizations at low SNRs.
	\item The sub-band method estimates the in-band spectral index values more accurately than MT-MFS, and the $\alpha$ values are reliable ($\Delta\alpha<0.2$) even for SNRs $\geq$ 50.  
	\item The broadband method has negligible bias compared to the uncertainties in the inferred $\alpha$ maps. The spectral index is reliable at SNRs as low as 15 as they produce an error of $\lesssim0.2$ in the absence of other systematics, like differences in the flux scale between bands and variations in the spectral index (e.g., spectral break, steepening, turnover) within the frequency range of interest.
	\end{enumerate}
For cases where the aim includes getting a reliable spectral index for mostly unresolved and faint (or a combination of bright and faint) source(s), observations using two different bands for broadband spectral indices will be the optimum use of the observing time. On the other hand, if the target source is bright enough (SNR $\gtrsim 100$), using an in-band spectral index may be optimum, particularly when improved uv-coverage for better image reconstruction may also be a requirement. For intermediate SNR, considering the added overhead time for observing in two different bands, it may be marginally beneficial to use the sub-band splitting method if the \textit{uv}-coverage is not a major requirement. Based on these, the user may optimize the observation and data analysis strategies depending on the required spectral index accuracy and the brightness of the targets. In our forthcoming work (Rashid et al., in prep.), we will extend this study to investigate further the reliability of spectral index on factors such as large-scale emission, wide-field effects, and their implications for uGMRT observations.
\par
Finally, we note that, for the Square Kilometre Array (SKA), precise spectral index measurements will be key requirement for many of the science goals - from understanding the details of star formation in nearby galaxies to accurate foreground removal for Epoch of Reionisation (EoR) studies using the redshifted 21-cm line \citep{2015aska.confE.171C,2017arXiv171206950A}. In this regard, new imaging algorithms, such as ``forced spectral fitting'' \citep{2023MNRAS.tmp.2352C}, with accurate spectral modeling can help in achieving improved deconvolution of sources and, in turn, better sensitivity limits. However, due to the enormous data volume for the SKA, there is discussion of the possibility that {\it only} selected data products (final image along with spectral index, polarization, and Faraday rotation maps), processed through a science data analysis pipeline, will be made available to the users \citep{2017ASPC..512..367H}. So, it will be of pivotal importance to systematically test the efficacy of the methods of spectral index estimation, similar to this exploratory study, for the SKA in the near future.

\begin{acknowledgments}
We thank the anonymous reviewer for useful suggestions and comments that helped us to revise and improve the manuscript. We are grateful to Urvashi Rao and Sanjay Bhatnagar for useful discussions and valuable input. This research has made use of NASA's astrophysics Data System. We acknowledge Max-Planck-Gesellschaft (MPG) for funding through the Max Planck Partner Group to carry out this research. 

\end{acknowledgments}

\vspace{5mm}

\software{CASA \citep{2007ASPC..376..127M}
        astropy \citep{astropy:2022},  
        SciPy \citep{2020SciPy-NMeth}, 
        NumPy  \citep{harris2020array}
          }\\
The codes to generate \textit{uv} data sets in this study can be found in the following repository: \url{https://github.com/rashid-astro/simuGMRT.git}.

\bibliography{spdex.bib}{}
\bibliographystyle{aasjournal}

\end{document}